\documentstyle [12pt]{article}

\title{Vacuum energy on orbifold factors of spheres}
\author{Peter Chang and J.S.Dowker %
\\{\em Dept.\ of Theoretical Physics, University of Manchester,} %
\\{\em Manchester M13 9PL, UK}}
\newcommand{\br}[1]{\mbox{\boldmath $#1$}}

\newcommand{\phv}{\varphi}

\newcommand{\pde}[1]{\partial_{#1}}

\newcommand{\inv}[1]{#1{-1}}
\newcommand{\shar}[3]{{\rm Y}{#1}_{#2}(#3)}
\newcommand{\legp}[2]{{\rm P}_{#1}(#2)}

\newcommand{\be}{\begin{equation}}
\newcommand{\ee}{\end{equation}}
\newcommand{\bea}{\begin{eqnarray}}
\newcommand{\eea}{\end{eqnarray}}
\newcommand{\bean}{\begin{eqnarray*}}
\newcommand{\eean}{\end{eqnarray*}}
\newcommand{\ba}[1]{\begin{array}{#1}}
\newcommand{\ea}{\end{array}}
\newcommand{\nno}{\nonumber}
\newcommand{\lef}[1]{\lefteqn{#1}}

\newcommand{\dd}{{\rm d}}
\newcommand{\half}[1]{\frac{#1}{2}}

\newcommand{\Tr}{{\rm Tr}}

\newcommand{\Real}{{\rm Re}}

\def\zf{$\zeta$-function}
\def\zfs{$\zeta$-functions}
\def\R{{\rm\rlap I\mkern3mu R}}
\def\Z{\rlap{\rm Z}\mkern3mu{\rm Z}}
\def\dalemb#1#2{{\vbox{\hrule height .#2pt
        \hbox{\vrule width.#2pt height#1pt \kern#1pt
                \vrule width.#2pt}
        \hrule height.#2pt}}}
\def\Sbox{\mathord{\dalemb{5.9}{6}\hbox{\hskip1pt}}}
\def\S{{\rm S}}
\def\H{{\rm H}}
\def\la{\lambda}
\def\si{\sigma}
\def\det{{\rm det\,}}
\begin{document}
\maketitle
\begin{abstract}
The vacuum energy is calculated for a free, conformally-coupled
scalar field on the orbifold space-time $\R\times \S^2/\Gamma$ where
$\Gamma$ is a finite subgroup of O(3) acting with fixed points. The energy
vanishes when $\Gamma$ is composed of pure rotations but not otherwise. It is
shown on general grounds that the same conclusion holds for all
even-dimensional factored spheres and the vacuum energies are given as
generalised Bernoulli functions. The relevant $\zeta$-functions are analysed
in some detail and several identities are incidentally derived. The general
discussion is presented in terms of finite reflection groups.
\end{abstract}
\thispagestyle{empty}
\clearpage
\pagenumbering{arabic}
\setcounter{page}{1}
\section{Introduction}

Roughly speaking an orbifold, or a V-manifold \cite{Sat}, is a quotient of a
manifold by a group of symmetries with fixed points \cite{Sco}. The orbifold is
singular at the images
of the fixed points where, for example, the Riemannian curvature would diverge.
Orbifolds can provide useful approximations to more complicated manifolds, such
as Calabi-Yau spaces, occurring in string compactifications \cite{Dix,Bagg,
Lanc}.
They have also appeared in membrane compactification \cite{Vas}. Some work has
also been done on geometric quantisation on orbifolds \cite{Mar}.

It is interesting to extend the usual theory of quantum fields on curved spaces
to orbifolds. Some cases have already been considered. The ideal cosmic string
is an example. We have earlier discussed some effects on the tetrahedron
\cite{DowT}.

In the present paper we consider a scalar field in the static space-time
$\R\times \S^2 \!/\Gamma$, where $\Gamma$ is one of the finite subgroups of
O(3) associated with the regular solids. The
fundamental domain of $\Gamma$ is a curvilinear triangle or quadrilateral
on $\S^2$ depending on whether $\Gamma$ does or does not include
reflections. The discussion is then extended to higher spheres.

As usual, we consider free fields and, in the present work, as an example
of a field theoretic calculation, we look at ways of evaluating the
conformal scalar contribution to the total Casimir energy on the divided
two-sphere. This will be sufficient to exhibit the various techniques and
relations.

Recently, there have been calculations of the vacuum energies
on the hyperbolic space-times, $\R\times \H^2 \!/\Gamma$, Zerbini et al
\cite{Zer}.

We have already shown that the total vacuum energy vanishes on the full sphere
$\R\times \S^2$ (and in fact on all even-dimensional spheres for conformal
coupling). It will turn out that dividing by a purely rotational $\Gamma$ does
not alter this conclusion. Including reflections does produce a nonzero
Casimir energy.

The analysis is presented in some calculational detail because we think the
discussion of the relevant heat-kernels and
$\zeta$-functions has some general methodological
value. The calculation can be thought of as a rather explicit example of an
equivariant situation, the general theory of which has been understood for a
long time.

\section{Vacuum energy}
We consider an ultrastatic spacetime, $\R\times M$, where the spatial section,
$M$, is $d$-dimensional.
A scalar field satisfies the traditional equation of motion
\be
(\Sbox + \xi R + m^2)\, \phv(x) = 0
\label{a1}
\ee
with the D'Alembertian $ \Sbox = \pde{t}^2 - \nabla^2 $.
$\nabla^2$ is the usual Laplace-Beltrami operator on $M$.

The general formula for the total vacuum, or Casimir, energy, $E$, for scalar
fields in a static space-time is given in the form we need in \cite{DnK}. Thus
\be
E ={1\over2}\lim_{s\rightarrow1}L^{-2(s-1)}\zeta_d(s-1/2)=
\half{1} \, \zeta_{d}({\textstyle -\half{1}}) \,,
\label{a9}
\ee
when this does not diverge, which it does not here. This will be discussed
later. $L$ is the scaling length.

The $\zeta$-function, $\zeta_{d}$, is a Dirichlet series for the operator
$(-\nabla^2 + \xi R + m^2)$ on $M$; $\zeta_d(s)=\sum_\la\la^{-s}$.
For constant scalar curvature, we can write
\be
\zeta_{d}(s) = \frac{1}{\Gamma(s)}
\int_0^\infty \dd \tau \, e^{- (m^2 + \xi R) \tau} \,\tau^{s-1} K_{d}(\tau)
\label{a10}
\ee
where $K_{d}$ is the integrated coincidence limit of the Laplacian heat-kernel
on $M$.

Incidentally, in the present case, the vacuum energy is also the negative of
the effective (one-loop) Lagrangian, $L^{(1)}$, which is related to the
effective
action, $W^{(1)}$, by
$W^{(1)} = \int \!\dd t \,L^{(1)}$.

\section{The Selberg trace formula on $\S^2\! / \Gamma$}
On the unit two-sphere, $\nabla^2\!\!=\!\!-{\bf L}^2$ where ${\bf L}$ is the
usual angular momentum operator. Then  the defining equation for the heat
kernel
on $\S^2$ is
\be
(\pde{\tau} + {\bf L}^2 )\, K_{\S^2} ({\br{r}},{\br{r}}';\tau)
= \delta(\tau)\, \delta({\br{r}},{\br{r}}')
\label{a11}
\ee
which has the solution
\bean
K_{\S^2} ({\br{r}},{\br{r}}';\tau) &=&
\sum_{l,m} e^{-l(l+1)\tau} \, \shar{l}{m}{{\br{r}}} \shar{l\,*}{m}{{\br{r}}'}
\\
&=& \frac{1}{4\pi}
\sum_{l=0}^{\infty} \,(2 l + 1)\, e^{-l(l+1)\tau} \,\legp{l}{\cos \Phi({\br{r}
},{\br{r}}')}
\eean
where $\cos \Phi({\br{r}},{\br{r}}') = \br{r} \cdot \br{r}'$, ${\br{r}}$ and
${\br{r}}'$ being unit three-vectors.

Next, the Mehler-Dirichlet integral representation for the Legendre
polynomial is used to get
\be
K_{\S^2} ({\br{r}},{\br{r}}';\tau) =
\frac{1}{2 \sqrt{2} \pi^2} \sum_{l=0}^{\infty}\, (2 l + 1)\, e^{-l(l+1)\tau}
\int_\Phi^\pi \dd \alpha\,
\frac{\sin (l+\half{1}) \alpha}{\sqrt{\cos \Phi - \cos \alpha}} \,,
\label{a12}
\ee
which is quite standard {\it e.g.} \cite{Per}.

We now restate our problem.
The spatial manifold we want to examine initially is the two-sphere with
a reduced symmetry, specifically $M \cong \S^2\! / \Gamma$ where
$\Gamma$ is a discrete subgroup of O(3). $\Gamma$ is either a cyclic
({\bf C}$_q$), dihedral
({\bf D}$_{2q}$), tetrahedral ({\bf T}), octahedral ({\bf O}) or icosahedral
group ({\bf Y}). If one uses an extended group, $\Gamma'$ ({\it i.e.}one that
includes a
reflection), then $\S^2/\Gamma'$ is a fundamental spherical triangle
in which the field satisfies Dirichlet or Neumann boundary conditions.
For simplicity, we restrict the analysis initially to pure rotational
$\Gamma\subset$ SO(3).

The method of images is used, with an argument similar to that found
in McKean and Lax \& Phillips \cite{McK,LnP}, to
find the heat kernel on the divided sphere. The idea is to write the
pre-image sum as a sum over conjugacy classes. It is easy to show that
\bea
K_{\S^2\! / \Gamma} (\tau) &=&
\int_{\S^2\! / \Gamma} \sum_{\gamma \in \Gamma}
K_{\S^2} ({\br{r}},\gamma {\br{r}};\tau)\, \dd {\br{r}} \nno \\
&=& \int_{\S^2\! / \Gamma} K_{\S^2} ({\br{r}},{\br{r}};\tau)\, \dd {\br{r}}
+ {1\over|\Gamma|}\sum_{\{\gamma\}}|\{\gamma\}|
\int_{\S^2}
K_{\S^2} ({\br{r}},\gamma {\br{r}};\tau)\, \dd {\br{r}} \,,
\label{a13}
\eea
where $\gamma$ is a typical element, one of $|\{\gamma\}|$, in the conjugacy
class $\{\gamma\}$. The elements of a class correspond to rotations
through one fixed angle about a set of conjugate axes. For a given set of such
axes, one corresponding class can be considered to be the primitive class, all
other classes
associated with these axes then being generated by this one. Thus the sum over
all classes can be rewritten as a sum over primitive classes and
the powers of these. Let $q$ be the generic order of the rotation
associated with the generic primitive class $\{\hat\gamma\}$ so that
$\hat\gamma^q={\rm id}$. Then $|\{\hat\gamma\}|$ is just the number,
$n_q$, of conjugate $q$-fold axes and we can write
\be
K_{\S^2\! / \Gamma} (\tau) = \int_{\S^2\! / \Gamma} K_{\S^2}
({\br{r}},{\br{r}};
\tau)\, \dd {\br{r}} + {1\over|\Gamma|}\sum_{\hat\gamma}n_q\sum_{p=1}^{q-1}
\int_{\S^2} K_{\S^2} ({\br{r}},{\hat\gamma}^p {\br{r}};\tau)\, \dd {\br{r}}.
\label{a14}
\ee
We can think of this rather obvious equation as a simple application of
Selberg's trace formula to a compact, here spherical, domain.
\section{Evaluation of the trace formula}
We will now evaluate each term in the above trace formula, (\ref{a14}).
The term corresponding to the identity element of the subgroup $\Gamma$,
from (\ref{a12}), is
\be
 \frac{1}{2 \pi^2} \sum_{l=0}^{\infty}\, (2l+1)\, e^{-l(l+1)\tau}
 \int_{\S^2\! / \Gamma} \dd {\br{r}}\,
 \int_0^{\frac{\pi}{2}} \dd \alpha\,\chi_l(2\alpha) ,
\label{a15}
\ee
where $\chi_l(2\theta)=\sin(2l+1)\theta/\sin\theta$ is the character of the
$l$-representation of SO(3).

Both integrals can be done to give
$$
 \frac{1}{|\Gamma|} \sum_{l=0}^{\infty}\, (2l+1)\, e^{-l(l+1)\tau}\,.
$$
Up to scalings, this last quantity is nothing other than the
extensively discussed rotational partition function of a gas of diatomic
molecules \cite{Mul,Per,Hur}.

However we do not want this form for the first trace term (\ref{a15}), instead
we shall leave it as follows
\be
 \frac{2}{\pi |\Gamma|} \sum_{l=0}^{\infty}\, (2l+1)\, e^{-l(l+1)\tau}
 \int_0^{\frac{\pi}{2}} \dd \alpha\, \chi_l(2\alpha)\,.
\label{a16}
\ee

Next we examine the second term in (\ref{a14}).
A little geometry shows that,
for $\gamma = \br{n}(\omega) \in\,${\rm SO}$(3)$,
$$
\cos \Phi({\br{r}},\gamma {\br{r}}) = \cos \omega
+ (1 - \cos \omega) (\br{r} \cdot \br{n})^2 \,.
$$
The summand of the double sum is then
\bea
\lef{ \frac{1}{2\sqrt{2} \pi^2}
 \sum_{l=0}^{\infty}\, (2l+1)\, e^{-l(l+1)\tau} } \nno \\
&& \mbox{} \times \int_{\S^2} \dd {\br {r}}\,
 \int_0^\pi \dd \alpha\, \sin (l+{\textstyle \half{1}})\alpha \,
\Real\, \frac{1}{\sqrt{c + (1-c) (\br{r}\cdot\br{n})^2 - \cos\alpha}}\,,
\label{a17}
\eea
where $c = \cos p\omega$. We may choose $\br{n}$ as the polar axis.

Pulling the space integral through, to change the order of integration, and
setting $\mu=\cos\theta$, (\ref{a17}) becomes
$$
 \frac{1}{\sqrt{2} \pi} \sum_{l=0}^{\infty}\, (2l+1)\, e^{-l(l+1)\tau}
 \int_0^\pi \dd \alpha\, \sin (l+{\textstyle \half{1}})\alpha
 \int_0^1 \!\dd \mu\,
 \Real\, \frac{2}{\sqrt{c - \cos\alpha + (1-c) \mu^2 }}\,.
$$
Since the radicand has to be positive, the two cases
of $c - \cos \alpha$ being either positive or negative must be considered.
Define the quantity $\overline{p\omega} = p\omega$ if $p\omega < \pi$
and $\overline{p\omega} = 2\pi - p\omega$ if $p\omega \geq \pi$.
Hence
\bean
\lef{ \frac{1}{\pi} \frac{\sqrt2}{\sqrt{1-c}}
\sum_{l=0}^{\infty}\, (2l+1)\, e^{-l(l+1)\tau} } \qquad\qquad\qquad \\
&\times& \left\{ \int_0^{\overline{p\omega}} \dd \alpha\,
\sin (l+{\textstyle \half{1}})\alpha
 \int_{\sqrt{Q(\alpha)}}^1 \dd \mu\,
 \frac{1}{\sqrt{\mu^2  - Q(\alpha)}} \right. \\
&&\left. \qquad\quad
\mbox{}+ \int_{\overline{p\omega}}^\pi \dd \alpha\,
\sin (l+{\textstyle \half{1}})\alpha
 \int_0^1 \dd \mu\,
 \frac{1}{\sqrt{\mu^2  + \bar{Q}(\alpha)}} \right\} \,,
\eean
where the terms in the braces can be evaluated to give
$$
\left\{
\int_0^{\overline{p\omega}} \dd \alpha\, \sin (l+{\textstyle \half{1}})\alpha
\inv{\cosh}\left( Q^{-\half{1}}(\alpha)\right)
+ \int_{\overline{p\omega}}^\pi \dd \alpha\,
\sin (l+{\textstyle \half{1}})\alpha
\inv{\sinh}\left( \bar{Q}^{-\half{1}}(\alpha)\right) \right\} \,.
$$
The functions $Q(\alpha)$ and $\bar{Q}(\alpha)$ are defined as
$Q(\alpha)=(\cos \alpha - c)/(1-c)$ and
$\bar{Q}(\alpha)=(c - \cos \alpha)/(1-c)$.

Integrating by parts leaves us with the desired, intermediate result,
\be
\frac{4}{\pi} \sum_{l=0}^{\infty}\, e^{-l(l+1)\tau}
P\!\!\int_0^{\frac{\pi}{2}} \dd \alpha\,
\frac{\cos (2l+1)\alpha\, \cos\alpha}{\cos 2\alpha - \cos p\omega} \,.
\label{pp}
\ee

The principal part, denoted by the $P$, is necessary in order to remove the
singularity at $\alpha= p\omega/2$ introduced by the partial integrations.

Finally, summing over the cyclic group, $\omega=2\pi/q$, (excluding the
identity, $p=q$), we get
\be
\frac{2}{\pi} \sum_{l=0}^{\infty}\, e^{-l(l+1)\tau}
P\!\!\int_0^{\frac{\pi}{2}} \dd \alpha\, \frac{\cos (2l+1)\alpha}{\sin\alpha}\,
\bigg[\cot\alpha - q \cot q \alpha\bigg]\,,
\label{a18}
\ee
using a standard  identity for the trigonometric sum.

Noting the Jacobi theta-function relation,
$$
e^{-\tau/4}\sum_{l=0}^{\infty}\,
e^{-l(l+1)\tau}\, \cos (2l+1)\alpha
= \sqrt{\frac{\pi}{4\tau}}
\sum_{n=-\infty}^{\infty} (-1)^n e^{-(\alpha + \pi n)^2\!/\tau} \,,
$$
and its derivative with respect to $\alpha$, we find the heat kernel, from
\ (\ref{a14}), (\ref{a16}) and (\ref{a18}), to be
\bean
\lef{ K_{\S^2 \!/ \Gamma} (\tau) = {2\pi\over|\Gamma|}\frac{e^{\tau/4}}
{(\pi\tau)^{3/2}}\sum_{n=-\infty}^{\infty} (-1)^n
\left\{\int_0^{\frac{\pi}{2}} \dd \alpha\,
(\alpha + \pi n) \frac{e^{-(\alpha+\pi n)^2\!/\tau}}{\sin \alpha} \right.
} \qquad\qquad\qquad\quad \\
&&\left.\mbox{}+{\tau\over2} \sum_{\hat{\gamma}} n_q P\!\!\int_0^{\frac{\pi}
{2}} \dd \alpha\,
\frac{e^{-(\alpha+\pi n)^2\!/\tau}}{\sin \alpha}\bigg[\cot\alpha - q \cot q
\alpha\bigg]\right\}\,.
\eean
It is possible to subsume the outer sum into the integration range,
producing the compact expression for the classical paths form of the trace
formula,
\be
K_{\S^2 \!/ \Gamma} (\tau) = {2\pi\over|\Gamma|}\frac{e^{\tau/4}}
{(\pi\tau)^{3/2}}P\!\!\int_0^{\infty} \dd \alpha\,
\frac{e^{-\alpha^2\!/\tau}}{\sin \alpha}\left\{\alpha
+{\tau\over2} \sum_{\hat{\gamma}}n_q \bigg[\cot\alpha - q \cot q \alpha\bigg]
\right\}\,.
\label{a19}
\ee

It should noted that this expression is akin to that obtained
in the hyperbolic case by Donnelly \cite{Don} and later by Balazs \& Voros
\cite{Bal}.

It is better to avoid the principal part by introducing a contour,
$C_\alpha$, located just above, or just below, the real $\alpha$ axis
and writing for (\ref{a19})

\be
 K_{\S^2 \!/ \Gamma} (\tau) = {\pi\over|\Gamma|}\frac{e^{\tau/4}}
{(\pi\tau)^{3/2}}\!\int_{C_\alpha} \dd \alpha\,
\frac{e^{-\alpha^2\!/\tau}}{\sin \alpha}\left\{\alpha
+{\tau\over2} \sum_{\hat{\gamma}}n_q\bigg[\cot\alpha - q \cot q \alpha\bigg]
\right\}\,.
\label{cf19}
\ee

By expanding the sines as infinite sums of fractions and running
$C_{\alpha}$ strictly along the real axis we can
confirm that this integral is equal to the principal part form appearing in
(\ref{a19}), \cite{DnC1}.

In the form  (\ref{cf19}) it is possible to effect a partial integration to get

\be
 K_{\S^2 \!/ \Gamma} (\tau) = {\pi\over|\Gamma|}\frac{e^{\tau/4}}
{(\pi\tau)^{3/2}}\!\!\int_{C_\alpha} \dd \alpha\,
\frac{e^{-\alpha^2\!/\tau}}{\sin \alpha}\left\{\alpha\big(1-\sum_
{\hat\gamma}n_q\big)-{\tau\over2} \sum_{\hat{\gamma}} n_qq \cot q \alpha\right
\}\,.
\label{cf20}
\ee
In the cyclic case, the term proportional to $\alpha$ goes away, leaving

\be
K_q (\tau) = -\frac{e^{\tau/4}}
{2(\pi\tau)^{1/2}}\int_{C_\alpha}\dd\alpha\,
\frac{e^{-\alpha^2\!/\tau}}{\sin \alpha}\cot q \alpha\,,
\label{ocp}
\ee
and (\ref{cf20}) can be written
\be
K_{\S^2 \!/ \Gamma} (\tau) =
{1\over|\Gamma|}\left[\sum_{\hat\gamma}qn_qK_q(\tau)-
(\sum_{\hat\gamma}n_q-1)K_1(\tau)\right]
\label{ghk}
\ee
where $K_1(\tau)=K_{\S^2}(\tau).$

There are several possible uses of these formulae, for example we could make an
asymptotic expansion to give a power series in the limit $\tau \rightarrow 0$,
\be
K_{\S^2/\Gamma}(\tau)\cong{1\over\tau}\sum_{k=0,1,\ldots}^\infty C_k\tau^k.
\label{asy}
\ee
This will be taken up later. The coefficients $C_k$ are related to values of
the $\zeta$-function given in the next section.
\section{The $\zeta$-function}

Important to the present work is a generalization of Riemann's zeta
function. The incomplete Riemann $\zeta$-function (also known as Hurwitz's
$\zeta$-function) is defined by
$$
\zeta_R(z,a) = \sum_{n=0}^{\infty} \frac{1}{(n+a)^z}
\,, \qquad \mbox{for } \Real\, z > 1\,.
$$
The analytic continuation of this, for all $z$, is
$$
\zeta_R(z,a) = -\frac{\Gamma(1-z)}{2\pi i}
\int_{C_0} \frac{(-x)^{z-1} e^{-a x} }{1 - e^{-x}} \dd x\,,
$$
where $C_0$ is the standard Riemann-Hankel contour around the positive real
axis \cite{WW}.
In particular we will have cause to use the incomplete $\zeta$-function
for the case $a=\half{1}$; so we give the contour representation
\be
\zeta_R(z,{\textstyle \half{1}}) = -\frac{2^{z-1} \Gamma(1-z)}{2\pi i}
\int_{C_0} \frac{(-x)^{z-1}}{\sinh x} \dd x\,.
\label{b1}
\ee

Turning now to the field equation (\ref{a1}), the case of conformal coupling
($m=0, \xi=1/8$) is particularly simple. Since $R=2$ for our scale, conformal
coupling is
equivalent to using the operator $({\bf L}^2+1/4)$ when finding the heat kernel
and amounts to dropping the $\exp(\tau/4)$ factor in (\ref{cf20}), which is the
simplifying fact.
Then the $\zeta$-function (\ref{a10}) is the Mellin transform of the integrated
heat kernel (\ref{cf20}), up to this factor. We have
\bea
\lef{ \zeta_{\S^2 \!/ \Gamma} (s) = \frac{1}{|\Gamma|\sqrt\pi\Gamma(s)}
\int_0^\infty \dd t\,t^{\half{1}-s}
\int_{C_\alpha} \dd \alpha\,
\frac{e^{-\alpha^2t}}{\sin \alpha}
} \qquad\qquad\qquad\quad \nno \\
&&\qquad\times
\left\{\alpha\big(1-\sum_{\hat\gamma}n_q\big) - {\tau\over2}\sum_{\hat{\gamma}}
qn_q\,\cot q \alpha
\right\}\,. \label{b2}
\eea
Note that we have set $t=1/\tau$. We will denote $\zeta_{\S^2 \!/ \Gamma}$
by $\zeta{{}_\Gamma}$.

The immediate objective is to invert the order of integration in (\ref{b2}) so
that the $t$-integral can be done.
Some manoeuvring is necessary before this can be achieved
because the $t$-integral will diverge when $\alpha$ is inside
the `light-cone' i.e. when $x^2\le y^2$, where $\alpha=x+iy$.
We always require ${\rm Re}(-\alpha^2 t)<0$ for convergence.

In terms of the real and imaginary parts, $\alpha = x + i y$ and $t = u + i v$,
$$
-\alpha^2 t = - (x^2 - y^2) u + 2 x y v - i [ (x^2 - y^2) v + 2 x y u ] \,.
$$

It is actually convenient to
take the average of two contour integrals, one, $C_\alpha$, above
the real $\alpha$-axis and the other, $C_\alpha'$, below. The $q=1$ integral
(corresponding to the full sphere) will then be

\be
\zeta_1(s)=\frac{1}{2\sqrt{\pi}\Gamma(s)}
\int_0^\infty \dd t \, t^{\half{1}-s}
\int_{C_\alpha+C_\alpha'} \dd \alpha\,
\frac{\alpha\,e^{-\alpha^2 t}}{\sin \alpha}\,.
\label{TC}
\ee
We study the behaviour as the $\alpha$- and $t$-contours are rotated.

Consider $C_\alpha$ rotated anti-clockwise onto the line $x=y$, then
the $t$-integral would be along the negative imaginary axis
of the complex $t$-plane.
Moving $C_\alpha$ further round to the vertical, $x=0$, would require the
$t$-contour, $C_t$, to run from the origin to $-\infty$ below the negative real
axis, which is a branch cut.
Similarly, rotating $C_\alpha'$ clockwise to the negative imaginary axis
of $\alpha$ implies that the corresponding contour $C_t'$ runs
above the branch cut from $t=0$ to $t=-\infty$.
As the $\alpha$-contours sweep passed the negative real axis, contributions
are picked up from the poles but these cancel and so can be neglected.
The next step is to combine the two $t$-contours, $C_t$ and $-C_{t}'$,
to form one continuous path, $C$, that loops clockwise round the negative
$t$-axis, passing through the origin. This is possible by reversing the rotated
$C_{\alpha}'$ contour, which runs parallel, but opposite, to the rotated
$C_{\alpha}$ one.

These two rotated $\alpha$-contours are next both moved onto the $y$-axis.
Redefining $\alpha\rightarrow i\alpha$ and using the symmetry of the integrand
in $\alpha$ we find,

\be
\zeta_1(s)=\frac{1}{i\sqrt{\pi}\Gamma(s)}
\int_C \dd t \, t^{\half{1}-s}
\int_0^\infty \dd \alpha\,
\frac{\alpha\,e^{\alpha^2 t}}{\sinh \alpha}\,.
\label{b5}
\ee

The inversion of the order of the $\alpha$- and $t$-integrations in (\ref{b5})
is effected
by deforming the $t$-contour, $C$, to run around the negative $t$-axis, {\it
avoiding the origin}. Then we get
$$\zeta_1(s)=\frac{1}{i\sqrt{\pi}\Gamma(s)}
\int_0^\infty \dd \alpha \,{\alpha\over\sinh \alpha}\,
\int_C \dd t\,t^{\half{1}-s}e^{\alpha^2 t}.$$
This should be compared with (\ref{TC}).

Integrating over $t$ yields
$$
\zeta_1(s)=-\frac{2\Gamma(\half{3}-s)}{\sqrt{\pi}\Gamma(s)}
\cos \pi s
\int_0^\infty \dd \alpha\,
\frac{\alpha^{2s-2}}{\sinh \alpha}\,,
$$
where we must assume that ${\rm Re}\, s>3/2$ in order that the
$\alpha$-integration converge at the origin.

Then, using (\ref{b1}), we finish with
$$
\zeta_1(s)=-\frac{\Gamma(\half{3}-s)\Gamma(2s-1)\cos \pi s}{\Gamma(s)2^{2s-3}
\sqrt{\pi}}
\zeta_R (2s-1,{\textstyle \half{1}})=2\zeta_R(2s-1,{\textstyle \half{1}}) \,.
$$
This is the conformal $\zeta$-function on the undivided sphere and is well
known, following immediately from the eigenvalues, $l(l+1)+1/4= (l+1/2)^2$ and
degeneracies $2(l+1/2)$,
\be
\zeta_{\S^2} (s) = 2 \zeta_R (2s-1, {\textstyle \half{1}}).
\label{b3}
\ee
(See Dowker [``Vacuum energies on spheres and in cubes'', unpublished 1983]
and Camporesi \cite{Cam}.)

For the other terms in (\ref{b2}), it can similarly
be shown that
$$
\frac{1}{\sqrt{\pi} \Gamma(s)}
\int_0^\infty \dd t\,
t^{-s-\half{1}}
\int_0^{\infty} \dd \alpha\,
\frac{e^{-\alpha^2t}}{\sin \alpha} \cot q \alpha
=
-\frac{2^{2s-1}}{\Gamma(2s)}
\int_0^{\infty} \dd \alpha\,
\frac{\alpha^{2s-1}}{\sinh \alpha} \coth q \alpha
$$
for $\Real\, s > 3/2$.
Then the entire $\zeta$-function,
(\ref{b2}), can be written,
\be
\zeta_{{}_{\S^2 \!/ \Gamma}}(s) =
{1\over|\Gamma|}\frac{2^{2s-1}}{\Gamma(2s)}
\int_0^{\infty} \dd \alpha\, \frac{\alpha^{2s-2}}{\sinh \alpha}
\left\{(2s-1)\big(1 - \sum_{\hat{\gamma}}n_q\big)
+\sum_{\hat{\gamma}}qn_q\coth q \alpha\right\}
\,, \label{b6}
\ee
valid for $\Real\, s > 3/2$. The same expression can be reached using the
principal part form (\ref{a19}).

In the case of the cyclic group, $\Gamma =\Z_q$, the first term in the braces
in the integrand of (\ref{b6}) vanishes and the result simplifies
to
\be
\zeta_{{}_{\S^2 \!/ Z_q}}(s)
=
\frac{2^{2s-1}}{\Gamma(2s)}
\int_0^{\infty} \dd \alpha\,
\frac{\alpha^{2s-1}}{\sinh \alpha} \coth q \alpha \,.
\label{b9}
\ee
A simpler, eigenvalue derivation of this formula is given in the next section.

Calling the $\zeta$-function of (\ref{b9})
$\zeta_q(s)$ for short, the general expression (\ref{b6}) reads
\be
\zeta_{{}_\Gamma}(s)={1\over|\Gamma|}\left[\sum_{\hat\gamma}qn_q\zeta_q(s)-
(\sum_{\hat\gamma}n_q-1)\zeta_1(s)\right]
\label{po}
\ee
where $\zeta_1(s)=\zeta_{{}_{\S^2}}(s)$ is given by (\ref{b3}).

For all $\Gamma$ except the cyclic groups, $\sum_{\hat\gamma}n_q-1=|\Gamma|/2$
and so we get
$$
\zeta_{\Gamma}(s)={1\over2}\zeta_{\S^2}(s)+{1\over|\Gamma|}\sum_{\hat\gamma}
qn_q\zeta_q(s)\,
$$
for $\Gamma\ne \Z_q$.

We now look for an analytic continuation of (\ref{b6}).
Expanding the hyperbolic cotangent in (\ref{b9}) in terms of exponential
functions and integrating, yields \cite[p361 (3.552.1)]{GR}
$$
\zeta_q(s)=
2\sum_{p=0}^{q-1} \sum_{m=0}^\infty
\frac{m}{\big(q m + p + \half{1}\big)^{2s}}
+ \zeta_R (2s,{\textstyle \half{1}}) .
$$
Rearranging the sums,
\be
\zeta_q(s)=
\zeta_R (2s,{\textstyle \half{1}})
+ \frac{2}{q} \zeta_R (2s-1,{\textstyle \half{1}})
- \frac{1}{q^{2s+1}}\sum_{p=0}^{q-1}
(2p+1) \zeta_R (2s,{\textstyle \frac{2p+1}{2q}})\,,
\label{b7}
\ee
which would serve as an analytic continuation using the Hurwitz
$\zeta$-function
relation \cite{WW} \cite[p1073 (9.521.2)]{GR}.

To produce a functional relation directly for the continuation of
$\zeta_q(s)$, we consider the integral
$$
-{2^{2s-1}\Gamma(1-2s)\over2\pi i}\int_{C_0} \dd \alpha\,
\frac{(-\alpha)^{2s-1}}{\sinh \alpha} \coth q \alpha.
$$
For $\Real\, s > 3/2$, this equals the integral in (\ref{b9}) and so is an
appropriate continuation.

For $\Real\, s < 0$, a large loop can be added without penalty to the contour
$C_0$ so forming a continuous path that ultimately surrounds all the poles on
the imaginary
axis. Then we get, from residues, and after rearranging the summations a
little,
$$
e^{2\pi is} \frac{\cos \pi s\Gamma(1-2s)}{q}
\left\{
\sum_{m \neq 0 \atop{\!\!\!\!\!\pmod{q}}} \frac{1}{\sin \frac{\pi m}{q}}
\left(\frac{\pi m}{q}\right)^{2s-1}
+
(2s-1)\sum_{m=1} (-1)^m
(\pi m)^{2s-2}
\right\} \,.
$$
Therefore, comparing with (\ref{b7}), the functional relation that must
exist is
\bea
\lef{
2^{1-2s} \Gamma(2s) \left\{
\zeta_R (2s,{\textstyle \half{1}})
- \frac{1}{q^{2s+1}}\sum_{p=0}^{q-1}
(2p+1) \zeta_R (2s,{\textstyle \frac{2p+1}{2q}})
\right\} } \nno
\\
&&\mbox{}=
\frac{\pi^{2s}}{q \sin \pi s}
\left\{
\sum_{p=1}^{q-1} \frac{1}{\sin \frac{\pi p}{q}}
\left[ \zeta_R(1-2s, {\textstyle \frac{p}{q}})
- 2^{2s} \zeta_R(1-2s, {\textstyle \frac{p+q}{2q}}) \right]
\right\} \,,
\label{b8}
\eea
which can be proved directly using the rational Hurwitz
$\zeta$-functional relation
and a couple of trigonometric sums found in Bromwich \cite[p272 Ex18]{Brom}.

{}From (\ref{b6}), (\ref{b7}) and (\ref{b8}), there results the explicit forms
\be
\zeta_{{}_\Gamma} (s) =
\frac{2}{|\Gamma|} \zeta_R (2s-1,{\textstyle \half{1}})
+{1\over|\Gamma|}\sum_{\hat{\gamma}}qn_q\left[
\zeta_R (2s,{\textstyle \half{1}})
- \frac{1}{q^{2s+1}}\sum_{p=0}^{q-1}
(2p+1) \zeta_R (2s,{\textstyle \frac{2p+1}{2q}})
\right]
\label{d1}
\ee
or, equivalently,
\bea
\lef{ \zeta_{{}_\Gamma}(s) =
\frac{2}{|\Gamma|} \zeta_R (2s-1,{\textstyle \half{1}})
} \label{d2} \\
&&
\mbox{} +
\frac{2^{2s}\Gamma(1-2s)\cos\pi s}{\pi^{1-2s}|\Gamma|}
\sum_{\hat{\gamma}}n_q\sum_{p=1}^{q-1}
\frac{1}{\sin \frac{\pi p}{q}}
\left[ \zeta_R(1-2s,{\textstyle \frac{p}{q}})
 - 2^{2s} \zeta_R(1-2s,{\textstyle \frac{p+q}{2q}}) \right]
\nno \,.
\eea

Equations (\ref{d1}) and (\ref{d2}) also exhibit the analytic properties
of the $\zeta$-function,
$\zeta_{{}_\Gamma}(s)$, which has only the one pole at
$s=1$ of residue $1/|\Gamma|$ in accordance with general theory, which gives
a residue of $|$S$^2/\Gamma|/4\pi$. Further, the coefficients in the asymptotic
expansion (\ref{asy}) are given by the values of the $\zeta$-function at
negative integers. The formula is
\be
\zeta_{{}_\Gamma}(-k)=(-1)^kk! C_{k+1}
\label{coeff}
\ee

It will be enough to give these values in the cyclic case,
\be
\zeta_q(-k)=-{1\over q(k+1)}B_{2k+2}(1/2)+
{q^{2k-1}\over2k+1}\sum_{p=0}^{q-1}(2p+1)B_{2k+1}\!\big((2p+1)/2q\big)
\label{ber}
\ee
in terms of Bernoulli polynomials.

The important value $\zeta_{{}_\Gamma}(0)$ is given in general by
\be
\zeta_{{}_\Gamma}(0)=C_{d/2}-n_0
\label{zet0}
\ee
where $n_0$ is
the number of zero modes. For conformal coupling $n_0$ vanishes. From
(\ref{d1}), or (\ref{b7}), we find ($d=2$)
$$
\zeta_q(0)={2q^2-1\over12q}.
$$

We also note, most easily from (\ref{d2}), that $\zeta_{{}_\Gamma}(s)$ has
zeros at $s=-(2k+1)/2$, with $k=0,1,\ldots$, in particular at $s=-1/2$.
Thus, from (\ref{a9}), the vacuum energy vanishes, as mentioned earlier. It
also vanishes on space-times with the structure
$\R\times\R^{2k}\!\times\S^2/\Gamma$. Of course, it must not be forgotten
that this conclusion depends upon the assumption of conformal coupling.

Expressions for the coefficients $C_k$ can be found by expansion of the
trigonometric functions in (\ref{a19}) or, more easily, in (\ref{ocp}) in
powers of $\alpha$. The result appears as a polynomial in $q$ whose
coefficients are products of Bernoulli numbers. Equivalence with
(\ref{ber}) devolves upon an interesting identity and is discussed in
Appendix A.

\section{A direct eigenvalue method}
In this section we describe the more direct eigenvalue method of deriving the
\zf and will later enlarge the discussion to include the extended groups. These
are easily allowed for by reinterpreting $\Gamma$ in (\ref{a13}) to be an
extended group and inserting phase (sign) factors in the sum such that the
direct elements, {\it i.e.} those with an even number of reflections, have a
factor of +1 while those elements with an odd number of reflections have a
factor of +1 or $-1$ depending, respectively, on whether we require Neumann or
Dirichlet boundary conditions at the edges of the fundamental domain.

On $\S^2 \!/\Gamma$, the eigenvalues of $({\bf L}^2+1/4)$
are $n^2/4$ for $n=2l+1$ and $l=0,1,2,\ldots$. If $d(l), \equiv d(n)$, is the
degeneracy of each eigenvalue labelled by $l$ then, by definition, the
conformal \zf on $\S^2/\Gamma$ is
\be
\zeta_{{}_\Gamma}(s)=2^{2s}\sum_{n=1,3,\ldots}{d(n)\over n^{2s}}.
\label{zetd}
\ee
{}From first principles $d(l)$ is the number of times the trivial
representation
of $\Gamma$ occurs in the $l$-representation of SO(3), ({\it cf} \cite{Stie})
\be
d(l)={1\over|\Gamma|}\sum_\gamma\chi_{{}_l}(\gamma)={2l+1\over|\Gamma|}+
{1\over|\Gamma|}\sum_{\hat\gamma}n_q\sum_{p=1}^{q-1}\chi_{{}_l}(\hat\gamma^p)
\label{degen}
\ee
where the cyclic character is
\be
\chi_{{}_l}(\hat\gamma^p)={\sin\big((2l+1)\pi p/q\big)\over\sin(\pi p/q)}\,.
\label{cchar}
\ee
Thus for $\Gamma=\Z_q$ (which is sufficient)
\be
d_q(l)={1\over q}\sum_{p=0}^{q-1}\chi_{{}_l}(\hat\gamma^p),
\label{char}
\ee
which reduces to $2[l/q]+1$, where $[l/q]$ is the integer part of $l/q$.

Evaluation of the \zf from (\ref{zetd}) proceeds as follows.
\bea
\zeta_q(s)&=&
{2^{2s}\over q}\sum_{p=0}^{q-1}{1\over\sin(\pi p/q)}
\sum_{n=1,3,\ldots}{\sin\big(n\pi p/q\big)\over n^{2s}} \nno\\
&=&{2^{2s}\over q}\sum_{p=0}^{q-1}{1\over\sin(\pi p/q)}
\sum_{n=1,2,\ldots}\left({\sin\big(n\pi p/q\big)\over n^{2s}}
-{\sin\big(2n\pi p/q\big)\over (2n)^{2s}}\right).
\label{zet}
\eea
Separating off the $p=0$ term and then setting $p\rightarrow q-p$ we find
\be
\zeta_q(s)={2\over q}\zeta_R(2s-1,1/2)+
{2^{2s}\over q}\sum_{p=1}^{q-1}{1\over\sin(\pi p/q)}
\sum_{n=1,2,\ldots}{\sin\big(n\pi p/q\big)\over n^{2s}}.
\label{zet2}
\ee
Next we recall Hurwitz's formula
\be
\sum_{n=1,2,\ldots}{\sin(2n\pi a)\over n^{2s}}={(2\pi)^{2s}\over4\Gamma(2s)
\sin\pi s}\left(\zeta_R(1-2s,a)-\zeta_R(1-2s,1-a)\right)
\label{Hurw}
\ee
to rewrite (\ref{zet2}) in terms of $\zeta_R$-functions,
\bea
\zeta_q(s)&=&{2\over q}\zeta_R(2s-1,1/2) \nno\\
&+&{(2\pi)^{2s}\over 4q\Gamma(2s)\sin\pi s}\sum_{p=1}^{q-1}{2^{2s}
\over\sin(\pi p/q)}\left[\zeta_R\big(1-2s,\textstyle{{p\over2q}}\big)-
\zeta_R\big(1-2s,\textstyle{{p+q\over2q}}\big)\right].
\label{zet3}
\eea

We could stop here but, if agreement with our previous
result, (\ref{d2}), is required, the relation,
$$\zeta_R(z,{\textstyle{p\over2q}})=2^z\zeta_R(z,{\textstyle{p\over q}})
-\zeta_R(z,{\textstyle{p+q\over2q}}),$$ can be used.

It is easy to introduce a U(1) twisting into the character formula,
(\ref{char}), by writing
\be
d_q(l,t)={2\over q}\sum_{p=0}^{q-1}{\sin\big((2l+1)\pi p/q\big)\over
\sin(\pi p/q)}\cos(2\pi pt/q)=d_q(l+t)+d_q(l-t)\,,
\label{char2}
\ee
where $t$ is an integer between 0 and $q$. One can think of this as the
effect
of an Aharonov-Bohm flux along the rotation axis. The modification will
clearly carry through unchanged
into the final formula, like (\ref{zet3}), for the \zf which as a
consequence will have the same zeros as before. The same conclusion holds
for the twisting,
\be
d(l)={1\over|\Gamma|}\sum_\gamma\chi_{{}_l}(\gamma)\chi^*_{{}_D}(\gamma)
={(2l+1)\dim D\over|\Gamma|}+{1\over|\Gamma|}\sum_{\hat\gamma}n_q\sum_{p=1}^
{q-1}\chi_{{}_l}(\hat\gamma^p)\chi^*_{{}_D}(\hat\gamma^p),
\label{degen2}
\ee
that results from taking the field to belong to any representation, $D$, of
$\Gamma$.

It must be said that there is no substantial difference between the
method of this section and that given in section 5. Not surprisingly, the
integral (\ref{pp}) can be done to yield
the expression (\ref{char}) for $d_q(l)$. We have taken the route we have in
order to exhibit various aspects of the analysis and to provide
checks of the formulae.

It is interesting therefore to make contact with the
work of Polya and Meyer \cite{Mey} who gave Molien generating functions
for the degeneracies of each harmonic basis of degree $l$ invariant under a
point group of three-dimensional Euclidean space. (See also Laporte
\cite{Lap}.)

The generating function is defined to be
\be
h(\si) = \sum_{l=0}^\infty d(l) \si^l \,.
\label{gfun}
\ee
{}From (\ref{char}), a geometric summation produces the cyclic form
$$h_q(\si)={(1+\si)\over q}\sum_{p=0}^{q-1}{1\over1-2\si\cos
(2\pi p/q)+\si^2},$$
showing the connection with the Molien determinant expression \cite{Mey},
A further summation yields, {\it cf} \cite{Mey} Table 1,
\be
h_q(\si) = \frac{1}{1-\si} \frac{1+\si^q}{1-\si^q}.
\label{me}
\ee
Expanding gives again $d_q(l)=2[l/q]+1$.

It is amusing to note that the generating function is, to a factor,
the integrated heat kernel of the (pseudo-) operator
$({\bf L}^2+1/4)^{1/2}$,
$$
K^{1/2}(\tau) \equiv
\Tr\left(\exp \big(-\tau \sqrt{{\bf L}^2+1/4}\big)\right)\,.
$$

This can be seen by setting $\si = e^{-\tau}$ so that, making redefinitions
where appropriate,
\bean
h(\tau) &=& \sum_l d(l)e^{-\tau l}=e^{\tau/2}\sum_{n=1,3,\ldots}d(n)e^{-\tau
n /2}  \\
&=& e^{\tau/2}
\Tr\left(\exp( -\tau \sqrt{{\bf L}^2+1/4})\,\right)
\eean
where the trace is over all eigenstates on $\S^2/\Gamma$.

The \zfs are related by
\be
\zeta(s) \equiv \zeta^{A}(s) \equiv \Tr\, A^{-s} =  \Tr\, (A^{1/2})^{-2s}
= \zeta^{A^{\half{1}}}(2s) \equiv \zeta^{1/2}(2s),
\label{c2}
\ee
with
$$
\zeta^{1/2}(s) = \frac{1}{\Gamma(s)}
\int_0^\infty \dd \tau\,\tau^{s-1}K^{1/2}(\tau)
$$
where, by the above,
$$
K^{1/2}(\tau) = e^{-\tau/2} h(\tau)\,.
$$
For the cyclic case using (\ref{me})
\be
K^{1/2}_q(\tau) = \frac{\coth q\tau/2}{2 \sinh \tau/2}\,.
\label{HK}
\ee
Therefore, for $\Real\, s > 3$, setting $\tau=2\alpha$,
\be
\zeta^{1/2}_q(s) = \frac{2^{s-1}}{\Gamma(s)}
\int_0^\infty \dd \alpha\, \frac{\alpha^{s-1}}{\sinh \alpha} \coth q\alpha\,,
\label{c3}
\ee
and, using (\ref{c2}), we regain the answer given above, (\ref{b9}), as we
must.

The general result (\ref{b6}) or (\ref{po}) also follows
using (\ref{degen}) and also is in accordance with the
equation on p.139 of \cite{Mey}. The $\zeta$-functions for the
finite subgroups of O(3) thus take the compact integral form, written using
the Hankel contour so as to hold for all $s$,
\be
\zeta_{{}_\Gamma}(s) =
\frac{i\Gamma(1-2s)}{2\pi} \int_{C_0} \dd \tau\,
(-\tau)^{2s-1}K^{1/2}_{{}_\Gamma}(\tau)
\,,
\label{intf}
\ee
where the $K^{1/2}_{{}_\Gamma}(\tau)$ are given by
\be
K^{1/2}_{{}_\Gamma}(\tau) =
{\cosh(d_0\tau/2)\over2\sinh (d_1\tau/2)\sinh(d_2\tau/2)} \,,
\label{form}
\ee
with the constants, $(d_0,d_1,d_2)$ equal to $(q,q,1)$ for $\Z_q$,
$(2q+1,2q,2)$
for {\bf D}$_{2q}$, $(6,4,3)$ for {\bf T}, $(9,6,4)$ for {\bf O} and
$(15,10,6)$ for {\bf Y}.

We have earlier shown that the $\zeta_{{}_\Gamma}(s)$ vanish at
$s=-(2k+1)/2$, $k=0,1,\ldots$. This can also be seen very simply as follows.

At $s=-(2k+1)/2$, the integrand of (\ref{intf}) is single-valued and
the contour can be shrunk down to a small loop around the origin
reducing the computation to one of residues.
The evenness of the integrand, (\ref{form}), means that the residue at the
origin is zero without the need for any particular representation.

In contrast, the \zfs for the {\it extended} groups do not vanish at
$s=-(2k+1)/2$. In the case $\R\times\S^2/\Gamma'$ we have explicitly
\be
K^{1/2}_{{}_{\Gamma'}}(\tau) =
{\exp({d_0\tau/2})\over4\sinh (d_1\tau/2)\sinh(d_2\tau/2)} \,,
\label{extk}
\ee
with the corresponding constants, $(d_1,d_2,d_0)$,
and find for the vacuum energy,
\be
E_{\Gamma'}=\pm{d_0\over96d_1d_2}\big(d_0^2-d_1^2-d_2^2\big)\,.
\label{vacen}
\ee
The sign depends on whether one chooses Dirichlet or Neumann conditions, which
itself is determined by setting the twisting, $\chi(A)$, equal either to
$\det A$ or to $1$, as described earlier.

The specific numerical values are $\pm11/192$ for ${\bf T}'$, $\pm29/256$ for
${\bf O}'$ and $\pm89/384$ for ${\bf Y}'$. (See Table 1.)
\section{Higher spheres and reflection groups}
We enlarge the analysis to space-times of the form
$\R\times\R^{2k}\times\S^d/\Gamma$
and also give a more systematic account of the extended groups.

In place of (\ref{a9}) there is now the general equation
\be
E={(-1)^k\over2(2\pi)^k(2k+1)!!}
\zeta_d(-k-1/2),
\label{nvac2}
\ee
when finite, for the vacuum energy per unit volume in the $\R^{2k}$ space.

We assume that, {\it for all} $k$, the scalar field is conformally coupled in
the static space-time, $\R\times \S^d/\Gamma$, so that the relevant (conformal)
eigenvalues are still $(n+d-2)/2$, $n=1,3,\ldots$.

In the present
case $\zeta_d$ is $\zeta_{{}_\Gamma}$ given by (\ref{intf})
where the heat kernel is related to the generating function by
\be
K^{1/2}(\tau) = e^{-(d-1)\tau/2} h(\tau)\,.
\label{hkl}
\ee

{}From (\ref{intf}) it follows that
$$\zeta_{{}_\Gamma}(-k-1/2)=-\left({\partial\over\partial\tau}
\right)^{2k+1}\!
e^{-(d-1)\tau/2}h(\tau)\bigg|_{\tau=0}.$$
Hence the behaviour of the combination $e^{-(d-1)\tau/2}h(\tau)=\si^{(d-1)/2}
h(\si)$ under $\tau\rightarrow-\tau$, {\it i.e.} under $\si
\rightarrow1/\si$, is relevant; if even, the residue is zero.

The structure of the (twisted) harmonic generating function, $h_\chi(\si)$, on
factored spheres, $\S^d/\Gamma$, is given by, \cite{Mey,BnH,Ber},
\be
h_\chi(\si)={(1-\si^2)\over|\Gamma|}\sum_{A}{\chi^*(A)\over\det(1-\si A)},
\label{hgf}
\ee
which incorporates Molien's theorem. The situation now is that $\Gamma$ is a
finite subgroup (extended or not, as the case may be) of O$(d+1)$ and the $A$
are its $(d+1)\times(d+1)$ matrix
representatives. A twisting, $\chi(A)$, has been included. $\chi(A)$ is
shorthand for $\chi(\gamma)$ where $A$ represents $\gamma$. Further, since $A$
is real orthogonal, $\chi
(A)$ could be replaced by its real part.

$h_\chi(\si)$ is a rational function but, at the moment, we
do not need to be more precise than (\ref{hgf}). Then it is easily shown that
\bea
\lefteqn{{\si}^{(d-1)/2}h_\chi(\si)-\si^{(1-d)/2}h_\chi(\si^{-1})=}\nno\\
&&{\si^{(d-1)/2}(1-\si^2)\over|\Gamma|}\sum_{A}{\chi^*(A)\big(1-(-1)^d\det
A\big)\over\det(1-\si A)}.
\label{diff}
\eea

For even-dimensional spheres, the sum in (\ref{diff}) reduces to one over those
elements of $\Gamma$ that contain an {\it odd} number of reflections
so immediately we see that $\si^{(d-1)/2}h_\chi(\si)$ is an even function when
$\Gamma$ contains only rotations. This shows quite generally that the \zfs on
such factored, even-dimensional spheres vanish at negative odd integers, (for
conformal coupling of course).
For odd spheres only the pure rotations contribute to (\ref{diff}).

The generating function (\ref{gfun}) is an example of a Poincar\'e (or Hilbert)
series and its derivation is an episode in Invariant Theory \cite{Spr}. For a
finite {\it reflection group}, which is another name for an extended group,
the structure of the usual harmonic generating function is just (\ref{hgf})
with $\chi(A)$ equal to unity,
\be
h_{{}_N}(\si)={(1-\si^2)\over|\Gamma|}\sum_A{1\over\det(1-\si A)}
=(1-\si^2)\prod_{i=1}^{d+1}{1\over(1-\si^{d_i})},
\label{refl}
\ee
where the $d_i$, $i=1,2,\ldots,d+1$, are the degrees of the algebraically
independent generating
members of the invariant polynomial ring associated with the action of the
group, $\Gamma$, on the $d+1$ dimensional vector space. Without loss of
generality, $\Gamma$ may be taken to be a subgroup of O($d+1$).

Since there is always
the basic invariant ${\br {r}}^2, =$ $x^2+y^2+z^2+\cdots$, of degree
$2,=d_{d+1}$, we see that the corresponding  term, $i=d+1$, in the denominator
of (\ref{refl}) is
cancelled by the  `harmonic factor', $(1-\si^2)$, leaving just the degrees
$(d_1,d_2,\ldots,d_d)\equiv{\bf d}$.

The properties of reflection groups
were worked out by Coxeter \cite{Cox}, Chevalley \cite{Che},
Shephard and Todd \cite{She}, Steinberg \cite{Stei} and Solomon \cite{Sol}.
 Flatto gives a readable review in \cite{Fla}.  An
introductory account can be found in Grove and Benson \cite{GnB}. For a more
advanced expository treatment see Humphreys \cite{Hum}. Physicists might care
to consult \cite{Mich}. Carter, \cite{Car},
contains useful algebraic information. Some standard, relevant facts follow.

For small $d$, the degrees $d_i$ can be determined by
using an identity relating them to the geometrical properties of the
group action. For example, it can be shown that $ \prod_i d_i$ is the order of
the group and that $\sum_i(d_i-1)$ equals the number of reflections. In the
two-sphere case these facts rapidly lead to the numbers cited earlier and we
see
that $d_0=d_1+d_2-1$ is the number of reflections.

Rewriting (\ref{refl}) in terms of the time $\tau$ we find
\be
h_{{}_N}(\tau)=e^{(d-1)\tau/2}e^{d_0\tau/2}\prod_{i=1}^d{1\over2\sinh(d_i
\tau/2)} .
\label{pgfnc}
\ee

Note that the effect
of the conformal coupling term, $e^{(d-1)\tau/2}$, is here associated with
the number of reflection hypersurfaces, the number of `positive roots'. Using
(\ref{hkl}) we can check that
(\ref{pgfnc}) reproduces the heat kernel (\ref{extk}) on $\S^2/\Gamma'$.

On restricting the reflection group to its rotational subgroup, another
generating
invariant polynomial comes into play. This is the Jacobian, of degree
$d_0=\sum_{i=1}^{d+1}(d_i-1)$, of the original generating set. Under the
action of
$\Gamma$, $J$ changes to $(\det A) J$. It is an odd form and, as such, is
the product of $d_0$ linear factors, the vanishing of each of which is the
equation of a reflecting hyperplane.

Multiplying invariant polynomials by $J$ produces {\it skew}-invariant
polynomials and conversely all such polynomials have this product form.
Thus their generating function is
\be
\prod_{i=1}^{d+1}{\si^{d_i-1}\over1-\si^{d_i}}=
\si^{d_0}\prod_{i=1}^{d+1}{1\over1-\si^{d_i}}={1\over|\Gamma|}\sum_A
{\det A\over\det(1-\si A)}.
\label{skgf}
\ee
This can be referred to as the {\it Dirichlet generating function} since the
odd elements of $\Gamma$ enter with a minus sign, {\it cf} \cite{Ber}. The
harmonic generating function is

\be
h_{{}_D}(\tau)=e^{(d-1)\tau/2}e^{-d_0\tau/2}\prod_{i=1}^d{1\over2
\sinh(d_i\tau/2)} ,
\label{pgfnc2}
\ee
The corresponding Neumann function has already been given in (\ref{pgfnc}).

Since skew-invariants are strictly invariant under pure rotations, the
generating function for $\Gamma\subset$ SO($d+1$) is
\be
(1+\si^{d_0})\prod_{i=1}^{d+1}{1\over1-\si^{d_i}}=e^{(d-1)\tau/2}
\cosh(d_0\tau/2)\prod_{i=1}^{d+1}{1\over2\sinh(d_i\tau/2)} .
\label{pgf}
\ee
Multiplying by the harmonic factor,
$(1-\si^2)=2e^{-\tau}\sinh\tau$, removes the $(d+1)$-th term from the
product and yields
\be
h(\tau)=h_{{}_N}(\tau)+h_{{}_D}(\tau)=2e^{(d-1)\tau/2}\cosh(d_0\tau/2)
\prod_{i=1}^d{1\over2\sinh(d_i\tau/2)} .
\label{phgf}
\ee

{}From (\ref{hkl}) and (\ref{phgf}) we can check (\ref{form}) for the heat
kernel on $\S^2/\Gamma$ in the rotational, SO(3) case.

Equation (\ref{intf}) with (\ref{extk}) is a particular case of Barnes'
\zf \cite{Bar} encountered in our earlier work \cite{Dow2}. We can write in the
$\S^2$ case
\be
\zeta_{{}_{\Gamma}}(s)=\zeta_2\left(2s,(d_1+d_2-d_0)/2\mid d_1,d_2\right)
=\zeta_2\left(2s,1/2\mid d_1,d_2\right).
\label{gbz}
\ee
The general definition of $\zeta_d(s,*\mid*)$ is
\bea
\zeta_d(s,a|{\bf {d}})&=&{i\Gamma(1-s)\over2\pi}\int_L d\tau {\exp(-a\tau)
(-\tau)^{s-1}\over\prod_{i=1}^d\big(1-\exp(-d_i\tau)\big)}\nno \\
&=&\sum_{{\bf {m}}={\bf 0}}^\infty{1\over(a+{\bf {m.d}})^s},\qquad
\Real\, s>d
\label{Barn}
\eea
where ${\bf {d}}=(d_1,d_2,\ldots,d_d)$ and ${\bf {m}}=(m_1,m_2,\ldots,m_d)$
with the $m_i$ non-negative integers.
The location of the infinite contour $L$ depends on the relative
positions of the numbers $a$ and ${\bf d}$, \cite{Bar}. In the present
case it can be taken to be the Hankel contour, $C_0$.

Barnes gives the values of $\zeta_d(s,*\mid*)$ at all integral values of $s$.
In
particular the values at the negative integers are the generalised
Bernoulli functions which result from a residue evaluation via an expansion of
the integrand. In particular
\be
\zeta_d(-q,a\mid{\bf d})={(-1)^dq!\over(d+q)!{\textstyle\prod}d_i}
B_{d+q}^{(d)}(a\mid{\bf d})
\label{ints}
\ee
in the notation of \cite{Erd}. Here, and in the following, the sums and
products run from $i=1$ to $i=d$.

The Bernoulli polynomials are developed most thoroughly by N\"orlund
\cite{Nor}.
As noted by Hirzebruch \cite{Hir} they are essentially the same as the
Todd polynomials.

In the general case, constructing the heat kernel one gets
\be
K^{1/2}_{\left({N\atop D}\right)}=e^{\pm d_0\tau/2}\prod{1\over2
\sinh(d_i\tau/2)} ,
\label{hklnd}
\ee

According to (\ref{hkl})
with (\ref{refl}), we find that the Neumann and Dirichlet \zfs on
$\S^d/\Gamma$ are given by (\ref{Barn}). Specifically,
\be
\zeta_{{}_N}(s)=\zeta_d\left(2s,(d-1)/2\mid {\bf d}\right),
\label{neu}
\ee
\be
\zeta_{{}_D}(s)=\zeta_d\left(2s,{\textstyle\sum}\,d_i-(d-1)/2\mid {\bf d}
\right),
\label{diri}
\ee
generalising the previous expression (\ref{gbz}).

We already know that the Neumann and Dirichlet vacuum energies on
$\R\times\R^{2k}\times\S^d/\Gamma$ are opposite in sign for even spheres
and of the same sign for odd ones. Equivalently, we could use
\be
B^{(m)}_n({\textstyle\sum}d_i-x\mid{\bf d})=(-1)^nB^{(m)}_n(x\mid{\bf d})
\label{bnf}
\ee
and the answer is
\be
E_{{}_{\Gamma}}= -(\mp)^{d+1}{(-1)^kk!\over|\Gamma|\pi^k
(d+2k+1)!} B^{(d)}_{d+2k+1}\left((d-1)/2\mid {\bf d}\right),
\ee
where we have used $2\prod d_i=|\Gamma|$.
Here, and later, the upper sign is for Neumann and the lower for Dirichlet
conditions and we recall that the field is conformally coupled in only
$(1+d)$-dimensions whatever the value of $k$.

The vacuum energy on $\R\times\R^{2k}\times\S^2/\Gamma$ is
$$E_{{}_{\S^2/\Gamma}}=\pm{(-1)^kk!
\over\pi^k(2k+3)!}B^{(2)}_{2k+3}\left(1/2\mid d_1,d_2\right).$$
For $k=0$, using
$$B^{(2)}_3(\gamma\mid\alpha,\beta)={1\over4}\left(2\gamma-\alpha-\beta\right)
\left(2\gamma(\gamma-\alpha-\beta)+\alpha\beta\right),$$
(\ref{vacen}) can be checked.

The rather trivial case of the one-sphere, {\it i.e.} the space-time
$\R\times\S^1/[q]$, gives the vacuum energy $E=-q/24$ for both Dirichlet and
Neumann conditions. For the circle ($q=1$), adding the two values produces
the standard periodic result, $E=-1/12$.

For the regular tessellations of the three-sphere, we find the values in Table
1.
\vskip 12truept
\begin{table}[htb]

\begin{center}
\begin{minipage}{8cm}
\begin{tabular}{|l|l|l|}    \hline
\hline
& & \\[-1.5ex]
\multicolumn{1}{|c|}{Group}    & \multicolumn{1}{c|}{Degrees}
& \multicolumn{1}{c|}{Casimir energy%
\footnote{Where there are two signs, the upper is for Neumann and
the lower for Dirichlet boundary conditions.}}  \\[0.5ex]
\multicolumn{1}{|c|}{$\Gamma$} & \multicolumn{1}{c|}{\bf d}
& \multicolumn{1}{c|}{$E$}\\[0.5ex] \hline
& & \\[-1.5ex] 
$[q]$      & $(q)$             &  $    -q / 24  $          \\[0.5ex]  \hline
& & \\[-1.5ex] 
$[3,3]$    & $(3,4)$           &  $\mp 11 / 192 $          \\[0.5ex]
$[3,4]$    & $(4,6)$           &  $\mp 29 / 256 $          \\[0.5ex]
$[3,5]$    & $(6,10)$          &  $\mp 89 / 384 $          \\[0.5ex]  \hline
& & \\[-1.5ex] 
$[3,3,3]$  & $ (3,4,5) $       &  $ -601 / 28800  $        \\[0.5ex]
$[3,3,4]$  & $ (4,6,8) $       &  $ -3557 / 46080 $        \\[0.5ex]
$[3,4,3]$  & $ (6,8,12)$       &  $ -69391 / 414720 $      \\[0.5ex]
$[3,3,5]$  & $ (12,20,30) $    &  $ -3178447 / 5184000 $   \\[0.5ex] \hline
\hline
\end{tabular}
\end{minipage}
\end{center}

\caption{Vacuum energies for the reflection groups on $S^1$,
$S^2$ and $S^3$.}
\label{tab1}
\end{table}
\section{The heat kernel expansion}
In place of (\ref{asy}) we have
\be
K_{\S^d/\Gamma}(\tau)\cong{1\over\tau^{d/2}}\sum_{k=0,1/2,\ldots}^
\infty C_k\tau^k,
\label{asyb}
\ee
where any half odd-integral powers are due to the boundary conditions.

Equation (\ref{coeff}) becomes
\be
\zeta_{{}_\Gamma}(-k)=(-1)^kk!C_{k+d/2}
\label{dcoeff}
\ee
and enables us to find $C_{k+d/2}$, $k$ an integer. A further result we need
is the following.

On a $d$-dimensional manifold with boundary, $\zeta_d(s)$ has poles at
$s=(d-m)/2$, for $m=0,1,\ldots,d-1$ and $m=d+1,d+3,\ldots$, with residues
\be
{1\over\Gamma\big((d-m)/2\big)}C_{m/2}.
\ee

The possibly nonzero heat kernel coefficients
are thus $C_0,C_{1/2}, C_{1},\dots,C_{d/2}$ and $C_{d/2+1},C_{d/2+2},\ldots$.

The importance of the coefficients, so far as field theory goes, is that in a
space-time of the form $\R\times\R^{2k}\times M_d$, $C_{(1+2k+d)/2}$ is
proportional to the coefficient of the pole term in
the effective Lagrangian and a nonzero value indicates an ultraviolet
divergence. Our interest here lies in even $2k$ (most typically zero),
since, even if no divergences appear, we do not wish to undertake the
technical evaluation of the derivative of the \zf that necessarily appears in
the vacuum energy. (The result can be expressed as a multiple
$\Gamma$-function,
\cite{Bar}.)

The issue of ultraviolet divergences thus amounts to the existence of a pole
at $s=-k-1/2$, as intimated in section 2. The Barnes
\zf (\ref{Barn}), $\zeta_d(s,a\mid{\bf d})$, has poles only at positive
$s,=1,2,
\ldots,d-1$, so there is no ultraviolet divergence and no conformal
anomaly. Expressions (\ref{a9}) and
(\ref{nvac2}) are thus sensibly finite, although, {\it de facto}, evaluation
is really enough, as we have seen.

We calculate the coefficients $C_{d/2+k}$, $k=1,2,\ldots$, using (\ref{dcoeff})
and (\ref{ints}).
{}From (\ref{intf}), the residue this time involves the even part of $K^{1/2}(
\tau)$ and we need
\bea
\lefteqn{{\si}^{(d-1)/2}h_\chi(\si)+\si^{(1-d)/2}h_\chi(\si^{-1})=}\nno\\
&&{\si^{(d-1)/2}(1-\si^2)\over|\Gamma|}\sum_{A}{\chi^*(A)\big(1+(-1)^d\det
A\big)\over\det(1-\si A)}.
\label{sum}
\eea
If $d$ is even, only pure rotations elements survive while, for odd
$d$, only odd parity ones remain; in particular the identity does not
contribute for odd $d$.

{}From (\ref{neu}) and (\ref{diri})
expressions for the coefficients in terms of generalised Bernoulli numbers can
be obtained
\be
C_{d/2+k}=(\mp)^{d}{(-1)^k2(2k)!\over|\Gamma|(d+2k)!k!}
B^{(d)}_{d+2k}\big((d-1)/2\mid{\bf d}),\quad (k=1,2,\ldots).
\label {coeff2}
\ee

The coefficient $C_{d/2}$ is given by (\ref{zet0})
\be
C_{d/2}=(\mp)^d{2\over|\Gamma|d!}B^{(d)}_d\big((d-1)/2\mid{\bf d}
\big).
\ee

The remaining coefficients can be found from the residues of the Barnes \zf
\be
C_{(d-k)/2}=(\mp)^{(d-k)}{2\Gamma(k/2)\over|\Gamma|
(k-1)!(d-k)!}B^{(d)}_{d-k}\big((d-1)/2\mid{\bf d}\big),\quad (k=1,2,\ldots,d).
\label {coeff3}
\ee

For odd-dimensional spheres, and purely rotational $\Gamma$, we thus
generalise the result that the heat kernel asymptotic expansion on the full
sphere terminates, in this case with the $C_{d/2-1/2}$ term.

A particularly interesting example is the
three-sphere since we can make contact with the results of Ray \cite{Ray}
on the relevant generating functions.

The asymptotic expansion of the heat kernel on orbifolds has been discussed by
Donnelly \cite{Don,Don2}, and Br\"uning \& Heintze \cite{BnH} amongst others.
The particular case of factored spheres is considered in some detail in
\cite{BnH}, the structure
of the Poincar\'e series being presented more abstractly than in \cite{Mey}.

The fact that the eigenvalues are proportional to squares of integers
means that (\ref{HK}) can be used immediately to give a series for the free
energy, $F$, of the finite temperature field theory. Thus, in general,
\be
F(\beta)=E+{1\over\beta}\sum_{m=1}^\infty {1\over m}K^{1/2}(m\beta)=E+
{1\over\beta}\sum_{m=1}^\infty{1\over m}
e^{\pm d_0m\beta/2}\prod{1\over2\sinh(d_im\beta/2)} ,
\ee
where $\beta=1/kT$.

In the cyclic case, on $\S^2$,
$$F_q(\beta)={1\over2\beta}
\sum_{m=1}^\infty {\coth(qm\beta/2)\over m\sinh(m\beta/2)}.$$
\section{Other couplings}
It is also possible to perform the calculation for massive or for minimally
coupled fields but then Bessel functions would arise and the analysis would be
more complicated({\it cf} \cite{Zer,Cam}).

The minimally coupled case is important as it corresponds to
conformal coupling on S$^2/\Gamma$ regarded as a Euclideanised space-time.
Interest would then centre on the evaluation of $\zeta'_{{}_\Gamma}(0)$ and of
$\zeta_{{}_\Gamma}(0)$, which is proportional to the conformal anomaly. In fact
it is easy to evaluate this last quantity using the results  of the present
paper since there is no extra difficulty in finding the heat kernel expansion
coefficients. The relation between the conformal heat kernel and that, $K_m$,
for the general equation of motion (\ref{a1}) is
\be
K_m(\tau)=\exp\tau\big(m^2+{d-1\over4}\big((4\xi-1)d+1\big)\big) K(\tau)
\ee
which, upon expansion, allows the new coefficients to be determined in terms of
the old in a well known way.

In the case of two dimensions, and minimal coupling, we find the conformal
anomaly
\be
\zeta_{{}_\Gamma}(0)=\big(C_1-{1\over4}C_0\big)={1\over
360|\Gamma|}\big(12(d_1+d_2)^2-4d_1d_2+60(d_1+d_2)-45\big).
\ee

The space-time region can be thought of as a curvilinear triangle, and the
conformal anomaly can also be found by conformal transformation methods.
\section{Conclusion}

We have found that, if $\Gamma$ contains only rotations, the total vacuum
energy, for conformal coupling on the divided even-dimensional sphere,
$\S^{2d}/\Gamma$, is the same as that on whole sphere, namely zero.
One would expect, however, that the vacuum average of the {\it local} energy
density, $\langle T_{00}(x)\rangle $, would be nonzero and would diverge as
the fixed points are approached, just as for a cosmic string.

If $\Gamma$ contains reflections, {\it i.e.} is an extended group, then the
vacuum energy is nonzero. In this case, because of the Dirichlet or Neumann
boundary conditions,  we would expect the local density to
diverge at the sides of the fundamental triangle.

The vacuum energy on an odd-dimensional sphere is always nonzero, since,
this time, it is the pure rotation elements of $\Gamma$ that contribute, and
these must be present. The curiosity here is that Neumann and Dirichlet
boundary conditions give the {\it same} result, just half that of the strictly
rotational $\Gamma\subset$ SO($d+1$) case, the half factor being simply a
volume effect.

A number of extensions suggest themselves. It would be straightforward to
evaluate the finite temperature
corrections and also to include the effect of a monopole magnetic field. In the
latter case, by adding, in an {\it ad hoc} fashion,
an extra term proportional to the square of the magnetic
field to the Hamiltonian, it is possible to make the eigenvalues perfect
squares so that the generating function can again be used. The only
difference is that the angular momentum quantum number starts at the monopole
number.

A further extension that might be considered is that to the Dirac equation.
However there appears to be an obstruction to setting up spin-half theory on
$\S^2/\Gamma$, at least using images.

\clearpage
\appendix \section{\bf A Bernoulli identity}

In the limit as $\tau \rightarrow 0$, the heat kernel can be expanded as a
power series:
$$
K_{\S^2\!/ \Gamma} (\tau) \approx
\frac{1}{\tau} \sum_{k=0}^{\infty} C_k \tau^k \,.
$$
We will just concentrate on the cyclic case, $\Gamma \cong \Z_q$.

As mentioned in section 5, the trigonometric functions in the integral
representation of the heat kernel, (\ref{a19}) or (\ref{ocp}) are written
as power series in $\alpha$ involving Bernoulli numbers, and the integration
then carried out. One finds, {\it cf} \cite{Bal},
\bean
\lef{K_q (\tau) = \frac{1}{q\tau}
\left\{ \sum_n (-1)^{n+1} (1-2^{1-2n})\,B_{2n} \,\frac{\tau^n}{n!} \right.
} \\
&& \left. \mbox{} +
2 \sum_{n,m} (-1)^{n+m+1} (1-2^{1-2n})\, \frac{B_{2n}\, B_{2m}}{(2n)!\, (2m)!}
\frac{(2n+2m-2)!}{(n+m+1)!}\, (1-q^{2m})\, \tau^{n+m} \right\} \,.
\eean
Manipulation of the double sum yields the heat kernel coefficients,
for $k \geq 0$,
$$
C_k = \frac{(-1)^{k+1}}{q\,k!} \left\{-B_{2k}({\textstyle \half{1}})
+
\frac{1}{2k-1} \sum_{m=0}^k {\textstyle {2k\choose 2m}} \,(q^{2m}-1)
B_{2k-2m}({\textstyle \half{1}})\, B_{2m}
\right\} \,.
$$
These coefficients give the values of the zeta function at negative integers,
via (\ref{coeff}).
The $C_k$ can also be found directly using the values of the Hurwitz zeta
function,
$\zeta_R(-k,b)= - B_{k+1}(b) / (k+1)$ for $k \geq 0$, and the result is given
in (\ref{ber}).

The relation needed to reconcile these two expressions is
the Bernoulli sum identity,
\be
\sum_{m=0}^k {\textstyle {2k\choose 2m}} \,(q^{2m}-1)
B_{2k-2m}({\textstyle \half{1}})\, B_{2m}
=
k\,q^{2k-2} \sum_{p=0}^{q-1}
(2p+1)\, B_{2k-1}({\textstyle \frac{2p+1}{2q}}) \,.
\label{bsum}
\ee
It is simple to verify this relation for $k = 0,1$ while for $k>1$ we have
recourse to Lemma 1 of Apostol, \cite{Apo}, which says that for $r > 1$, there
is the identity,
\be
2r\sum_{\mu=1}^{\nu-1} {\textstyle \frac{\mu}{\nu}}\,
B_{2r-1}({\textstyle \frac{\mu}{\nu}})
 = \sum_{t=0}^{2r}
{\textstyle {2r\choose t}}\, B_t\, B_{2r-t}\, \nu^{1-t}
+ (2r-1)\, \nu^{1-2r} B_{2r}.
\label{bern}
\ee
This can be used to get
\bean
\sum_{p=0}^{q-1}\,(2p+1)\,B_m( {\textstyle \frac{2p+1}{2q} })
&=& \frac{q}{2k}\sum_{m=0}^k
{\textstyle {2k\choose 2m}}\, B_{2m}\, B_{2k-2m}\,
q^{1-2m} ( 2^{2-2m} - 2 ) \\
&& \mbox{} +
\frac{2k-1}{2k} \, q^{2-2k} ( 2^{2-2k} - 2 )\, B_{2k} \,.
\eean
Thus
$$
k\,q^{2k-2} \sum_{p=0}^{q-1}
(2p+1)\, B_{2k-1}({\textstyle \frac{2p+1}{2q}})
= (2k-1) B_{2k}({\textstyle \half{1}})
+ \sum_{m=0}^k
{\textstyle {2k\choose 2m}}\, q^{2k-2m} B_{2k-2m}\,
B_{2m}({\textstyle \half{1}})\,,
$$
and we are left with having to show that
\be
(2k-1) B_{2k}({\textstyle \half{1}}) =
- \sum_{m=0}^k {\textstyle {2k\choose 2m}} \,
B_{2k-2m}({\textstyle \half{1}})\, B_{2m}
\equiv \sum_{m=0}^{2k} {\textstyle {2k\choose m}} \,
(1 - 2^{1-m})\, B_{2k-m}\, B_{m} \,.
\label{proo}
\ee
Again the identity (\ref{bern}) from Apostol, \cite{Apo}, gives
$$
\left(2k B_{2k-1} - (2k-1)B_{2k}\right) -
2 \left( \frac{2k}{4} B_{2k-1} ({\textstyle \half{1}})
-\frac{2k-1}{2^{2k}} B_{2k}\right)
$$
for the right hand side of (\ref{proo}). This simplifies to
$$(2k-1) (2^{1-2k}-1) B_{2k}$$
and so  we have proved the Bernoulli sum, (\ref{bsum}).

\enddocument